\documentclass[final,aps,prb,preprint,superscriptaddress,showpacs]{revtex4-1}

\usepackage{graphicx}
\usepackage{dcolumn}
\usepackage{amsmath,bm}
\usepackage{color}%
\usepackage{multirow}
\usepackage{array} 
\usepackage{esvect}
\usepackage{braket}
\usepackage{makecell}
\usepackage[utf8]{inputenc}
\usepackage{kotex}
\usepackage{xr}
\usepackage{soul,xcolor}
\setstcolor{red} 

\usepackage{xr}

\makeatletter
\newcommand*{\addFileDependency}[1]{
  \typeout{(#1)}
  \@addtofilelist{#1}
  \IfFileExists{#1}{}{\typeout{No file #1.}}
}
\makeatother

\newcommand*{\myexternaldocument}[1]{%
    \externaldocument{#1}%
    \addFileDependency{#1.tex}%
    \addFileDependency{#1.aux}%
}

\myexternaldocument{SI}  

\begin{document}

\title{Orientation disparity in GaN/graphene/$m$-sapphire: control-based re-examination of thru-hole epitaxy} 

\author{Su Young An}
\thanks{these authors contributed equally}
\affiliation{Department of Physics and Research Institute for Basic Sciences, Kyung Hee University, Seoul, 02447, Republic of Korea}

\author{Hyunkyu Lee}
\altaffiliation{these authors contributed equally, Current address: 1.Center for 2D Quantum Heterostructures, Institute for Basic Science (IBS), Suwon 16419, Republic of Korea, 2.Sungkyunkwan University (SKKU), Suwon 16419, Republic of Korea}
\affiliation{Department of Information Display, Kyung Hee University, Seoul, 02447, Republic of Korea}

\author{Gunhoon Beak}
\author{Hyeonoh Jo}
\author{Jae Hun Kim}
\author{Jongwoo Ha}
\affiliation{Department of Physics and Research Institute for Basic Sciences, Kyung Hee University, Seoul, 02447, Republic of Korea}

\author{Jieun Yang}
\affiliation{Department of Chemistry, Kyung Hee University, Seoul, 02447, Republic of Korea}

\author{Changwook Dong}
\author{Jaewu Choi}
\author{Joonwon Lim}
\affiliation{Department of Information Display, Kyung Hee University, Seoul, 02447, Republic of Korea}

\author{Chinkyo Kim}
\email{ckim@khu.ac.kr}
\affiliation{Department of Physics and Research Institute for Basic Sciences, Kyung Hee University, Seoul, 02447, Republic of Korea}
\affiliation{Department of Information Display, Kyung Hee University, Seoul, 02447, Republic of Korea}

\begin{abstract}
The crystallographic orientation of films grown on 2D-masked substrates is often used to infer the pathway among remote, van der Waals, and thru-hole (pinhole-seeded) epitaxy.  However, attribution of a specific growth mechanism based on orientation can be ambiguous unless \emph{mask continuity} and \emph{substrate pre-treatment} are evaluated within a single process window. We compare GaN grown under identical conditions on four $m$-plane sapphire templates: (i) bare, (ii) “graphene-grown” (high-temperature Ar/H$_2$ with CH$_4$ on), (iii) “anneal-only” (high-temperature Ar/H$_2$ with CH$_4$ off), and (iv) graphene oxide spin-coated and reduced on pristine sapphire. GaN selects $(103)$ on graphene-grown and anneal-only $m$-plane sapphire, selects $(100)$ on bare $m$-plane sapphire, and is \emph{predominantly} $(100)$ with a \emph{minority} $(103)$ on graphene oxide spin-coated and reduced/pristine $m$-plane sapphire. High-resolution TEM shows that, on partly graphene-covered samples, nucleation occurs on \emph{exposed sapphire} (thru-hole), not on graphene, providing mechanism evidence independent of orientation. Within this window, the \emph{substrate surface state} set by high-temperature Ar/H$_2$ pre-treatment—rather than mask continuity—primarily governs orientation, while open-area effects can play a secondary role. Thus, \emph{preferred orientation alone may not determine the growth mechanism}; mask continuity and substrate pre-treatment must be explicitly controlled when using orientation as evidence for mechanism assignment.
\end{abstract}
\maketitle

\section{Introduction}

Remote epitaxy (RE) and thru-hole (pinhole-seeded lateral) epitaxy are best treated as \emph{regime-dependent} rather than mutually exclusive pathways. When a transferred or grown 2D layer is sufficiently thin and continuous (no openings), crystallographic registry through the 2D layer is commonly interpreted within an RE framework.\cite{Kong-NM-17-999} When openings are present, nucleation on exposed substrate areas followed by lateral overgrowth across the 2D layer is consistent with thru-hole (pinhole-seeded lateral) epitaxy.\cite{Lee-CGD-22-6995,Jang-AMI-10-2201406,Manzo-NC-13-4014} In practice, interfaces may fall between these limits; the pathway depends on \emph{mask continuity}, \emph{substrate/2D pre-treatment}, and the specific growth window; direct comparison requires evaluating these controls within one process window.

Within this broader picture, it is useful to recall that on $m$-plane sapphire the \emph{preferred orientation} of GaN is sensitive to substrate conditioning and buffer design. Adjusting the pretreatment/buffer/temperature window can switch the dominant orientation among $(10\bar{1}0)$, $(1\bar{1}00)$, and $(11\bar{2}2)$.\cite{Jue-SR-5-16236} In particular, AlN buffer engineering on $m$-plane sapphire steers $m$-GaN growth characteristics and orientation selection, underscoring the role of the initial interface and surface state.\cite{Paduano-JCG-367-104} Related nitridation/buffer studies on $m$-plane sapphire show that high-temperature pretreatments bias orientation outcomes (e.g., single $(10\bar{1}3)$ AlN under specific conditions), consistent with substrate-chemistry and step-morphology control of nucleation.\cite{Zhang-Micromachines-12-1153} In parallel, work on graphene/sapphire (2D-masked epitaxy) indicates that thermal annealing of transferred-graphene interfaces and other process parameters can shift the balance among \emph{remote}, \emph{pinhole}, and \emph{van der Waals} pathways.\cite{Du-NL-22-8647} These observations motivate explicit controls that separate \emph{mask continuity} from \emph{substrate modification} within a single process window.

A recent study on GdPtSb employed graphene with openings on sapphire and reported a preferred-orientation difference relative to growth on bare sapphire under identical conditions.\cite{Du-NL-22-8647} Because nucleation through openings could, in principle, mimic growth on bare sapphire, the authors argued against PSLE and considered RE as the likely pathway. That inference hinges on an assumption: nucleation on exposed sapphire should select the same preferred orientation as growth on fully bare sapphire under the same conditions.

Here we test that assumption using GaN on $m$-plane sapphire within a single process window. We compare four templates grown under identical GaN conditions: (i) bare $m$-plane sapphire, (ii) “graphene-grown” $m$-plane sapphire (high-$T$ Ar/H$_2$ with CH$_4$ on), (iii) “anneal-only” $m$-plane sapphire (high-$T$ Ar/H$_2$ with CH$_4$ off), and (iv) graphene oxide (GO) spin-coated and reduced on bare $m$-plane sapphire. High-resolution TEM on partly graphene-covered samples identifies the nucleation locus on \emph{exposed sapphire} (thru-hole), not on graphene. Despite this common locus, the preferred orientation differs across templates. Thus, orientation differences can arise within the thru-hole regime and do not by themselves imply RE.

These results lead to a practical point for regime assignment: orientation alone does not determine mechanism. Controls that separate \emph{mask presence/continuity} from \emph{substrate pre-treatment} within the same window help avoid misattribution. We adopt that approach below and examine how pre-treatment associated with CVD-growth of graphene can modify the $m$-plane sapphire surface in ways that affect orientation selection, even when nucleation occurs at exposed areas.  Throughout, we use “graphene with openings” to denote graphene layers that contain thru-holes or pinholes exposing the underlying sapphire (open-area fraction).


\section{Experimental}
Graphene was grown directly on $m$-plane sapphire by chemical vapor deposition at $1050^{\circ}$C and 400~Torr, with gas flows of Ar/CH$_4$/H$_2$ = 600/15/10~sccm for 3~h. The bare $m$-plane sapphire and the graphene-grown $m$-plane sapphire are denoted S1 and S2, respectively. A third $m$-plane sapphire substrate underwent the same thermal recipe as that of S2 with CH$_4$ off (“anneal-only”) and is denoted S3. For S4, an rGO film was fabricated by spin-coating the tetrabutylammonium (TBA)-GO solution followed by thermal reduction.\cite{Beak-CGD-25-7557} The GO solution was prepared by dispersing graphite oxide, synthesized via the Hummers method, in deionized (DI) water, followed by exfoliation using a sonicator. Large, unexfoliated particles were removed through centrifugation, and residual ions were eliminated through a dialysis process to obtain a high-purity GO solution.  The concentration of the TBA-GO solution was adjusted to 0.5 mg/mL, and spin-coating was performed at 3000 rpm for 1 minute. After spin-coating, thermal reduction was carried out at 890$^\circ$C for 30 minutes to convert graphene oxide (GO) into reduced graphene oxide (rGO). GaN was then grown on S1--S4 by hydride vapor phase epitaxy at $945^{\circ}$C, with NH$_3$ and HCl flows of 600 and 4~sccm, respectively, for 3~min.

In this study, GaN growth was carried out without the use of a low-temperature buffer layer, which is typically employed to achieve a continuous and flat GaN film. Instead, GaN was grown in the form of isolated domains, allowing for a direct assessment of the preferred orientation of these domains through visual inspection using scanning electron microscopy.\cite{Lee-CGD-22-6995} Additionally, X-ray diffraction analysis was performed to confirm the preferred orientation of GaN domains, whether they were grown on graphene/$m$-plane sapphire or on bare $m$-plane sapphire surfaces.

\section{Results and discussion}

\subsection{Bare and CVD-grown-graphene-covered $m$-plane sapphire}
To conduct a comparative investigation into the impact of a partially covering graphene substrate, serving as a mask layer, on the crystallographic orientation of GaN when grown on $m$-plane sapphire, we prepared a graphene/$m$-plane sapphire substrate by directly growing graphene on the $m$-plane sapphire substrate.

\begin{figure}
\includegraphics[width=0.7\columnwidth]{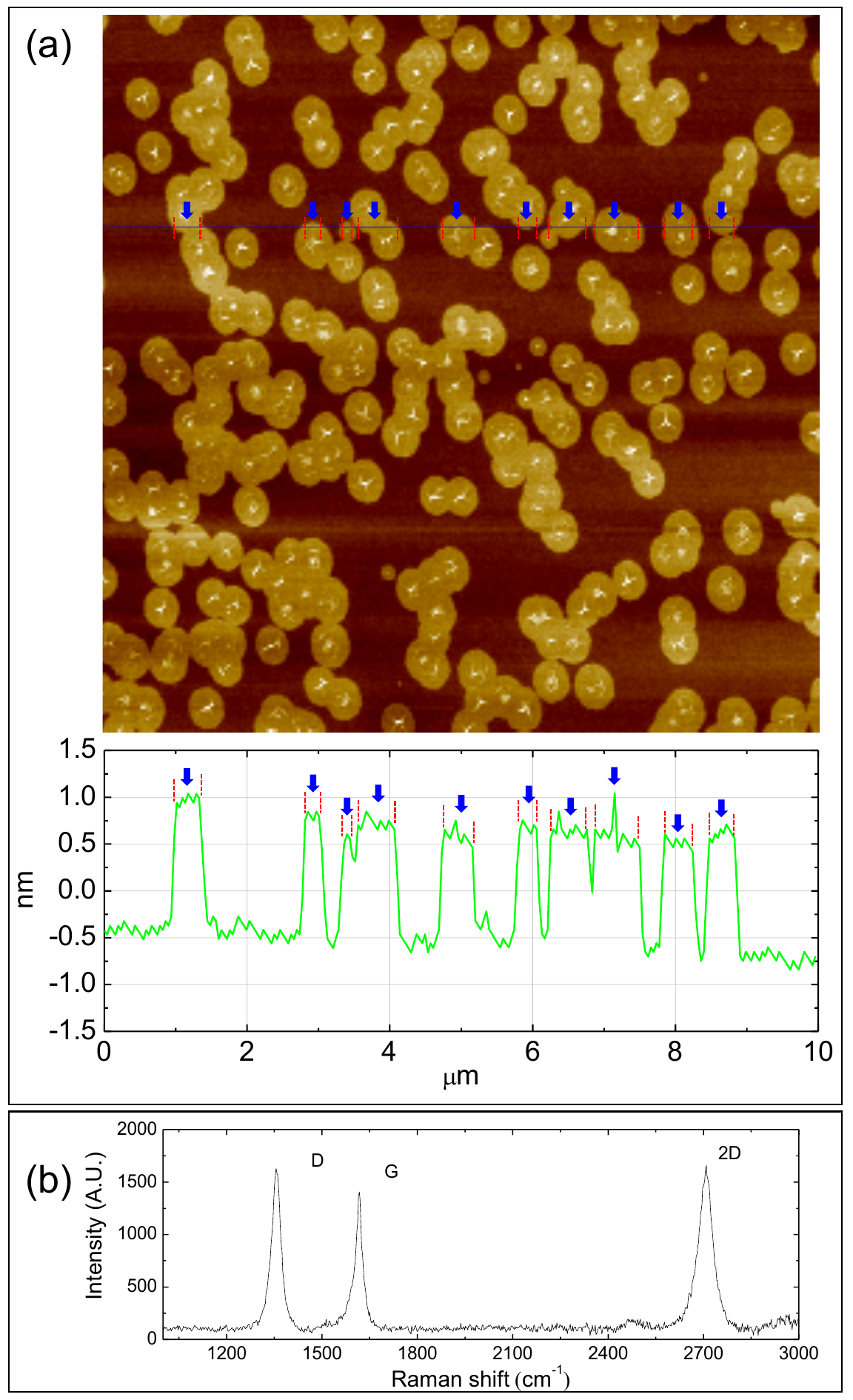}
\caption{(a) AFM topography of graphene directly grown on an $m$-plane sapphire substrate. Shown below is a height profile of graphene/$m$-plane sapphire along the blue line. (b) Raman spectroscopy of graphene grown on $m$-plane sapphire substrate S2.    
}
\label{graphene}
\end{figure}

In Fig.\ref{graphene} (a), we present the AFM topography along with a line profile of the graphene directly grown on an $m$-plane sapphire substrate. Here, approximately one-nanometer-thick round-shaped domains are densely distributed across the substrate, resulting in an estimated areal coverage of graphene of approximately 45\%, as determined by ImageJ analysis. Notably, the graphene domains only partly cover the $m$-plane sapphire substrate. We tuned the open-area fraction so that exposed sapphire was available for nucleation. Consequently, the areal coverage of graphene in our sample is smaller than that of graphene with openings as utilized in the previous study\cite{Du-NL-22-8647}. Raman spectra, as depicted in Fig.\ref{graphene} (b), confirm that these domains consist of a few-layer-thick graphene\cite{Hwangbo-Carbon-77-454}, consistent with the thickness measurement obtained from the AFM line profile.

\begin{figure}
\includegraphics[width=0.55\columnwidth]{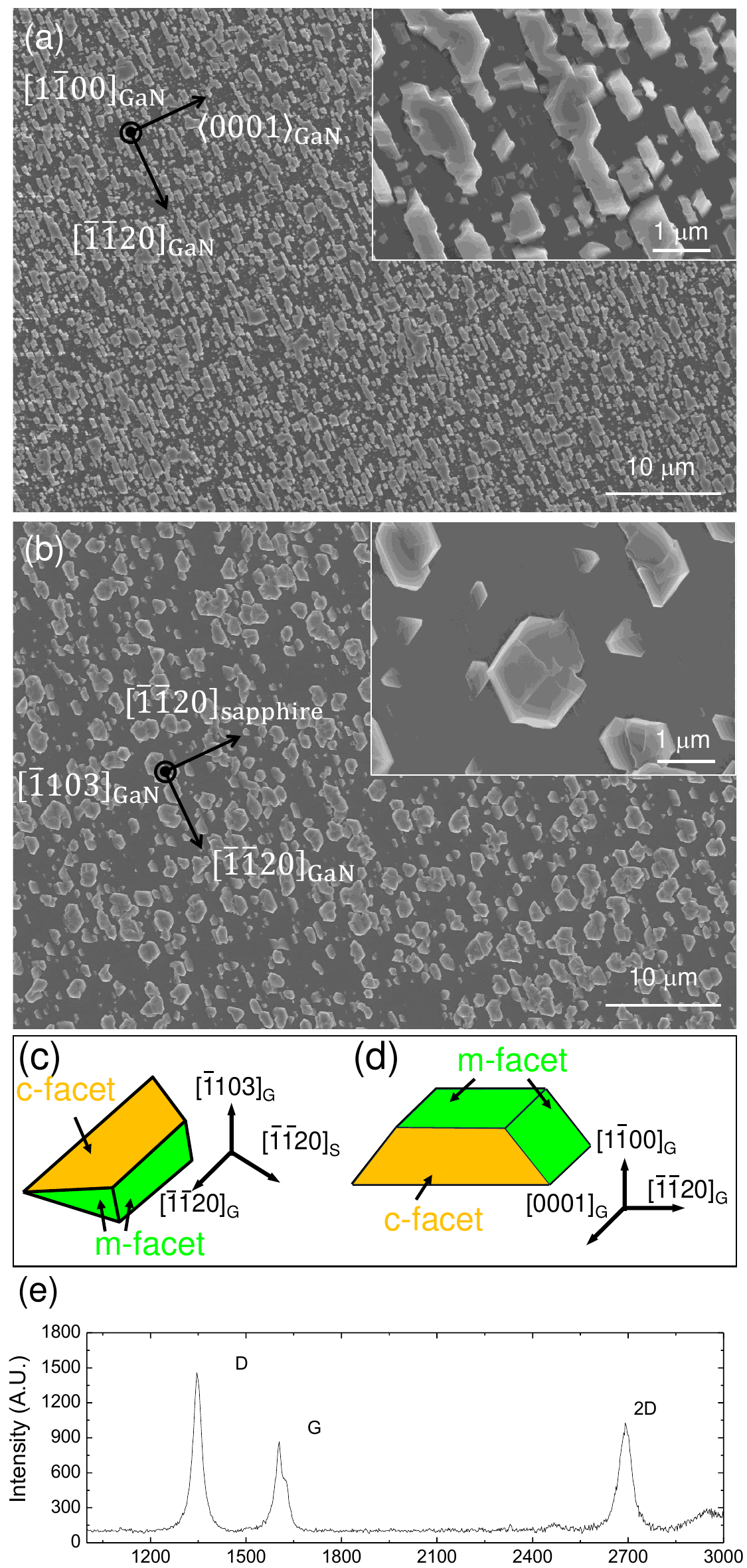}
\caption{SEM images of GaN domains grown on (a) a bare $m$-plane sapphire or (b) a partly graphene-covered $m$-plane sapphire substrate.  The characteristic shapes of GaN domains on substrate S1 and S2 are associated with (100)- and (103)-oriented domains, respectively. Schematic diagrams of (c) (103)-oriented and (d) (100)-oriented GaN domains. (e) Raman spectroscopy of graphene after GaN was grown on a partly graphene-covered $m$-plane sapphire substrate (S2).
}
\label{SEM-dissimilar-preferred-orientations-evidence}
\end{figure}

Two $m$-plane sapphire substrates were prepared, with one (S1) left bare and the other (S2) having graphene partly grown on it. Both substrates were subjected to GaN growth under identical conditions. As illustrated in Fig.~\ref{SEM-dissimilar-preferred-orientations-evidence}, scanning electron microscopy (SEM) images were acquired for each of the two samples. Given that a low-temperature buffer layer was not employed, GaN was grown as isolated islands with well-defined facets rather than forming a continuous film.  SEM shows that the preferred orientation of the GaN domains grown on the two substrates was dissimilar. To be more specific, the domain shapes in Fig.\ref{SEM-dissimilar-preferred-orientations-evidence}(a) and (b) exhibited characteristic features associated with (100)- and (103)-oriented GaN domains as shown in (c) and (d), respectively.\cite{Vennegues-JAP-108-113521,Jang-AMI-10-2201406,Jue-APL-104-092110,Seo-APEX-5-121001,Jue-SR-5-16236,Yoon-JAC-48-195,Lee-APL-104-182105} A more rigorous analysis of the crystallographic orientation of the grown GaN domains is given later using X-ray diffraction.

To ascertain whether the graphene layer, directly grown on $m$-plane sapphire, remained intact during the subsequent GaN growth on substrate S2, we conducted Raman spectroscopy measurements. The Raman measurement in fig.~\ref{SEM-dissimilar-preferred-orientations-evidence}(e) revealed that the graphene remained intact, exhibiting no significant damage during the GaN growth process. This observation aligns with previously reported findings regarding the stability of graphene on sapphire substrates while the damage of graphene on III-nitrides was observed at temperatures above the nitride decomposition threshold. \cite{Park-AMI-6-1900821,Lee-JAC-53-1502,Park-CS-12-7713}

\begin{figure}
\includegraphics[width=1.0\columnwidth]{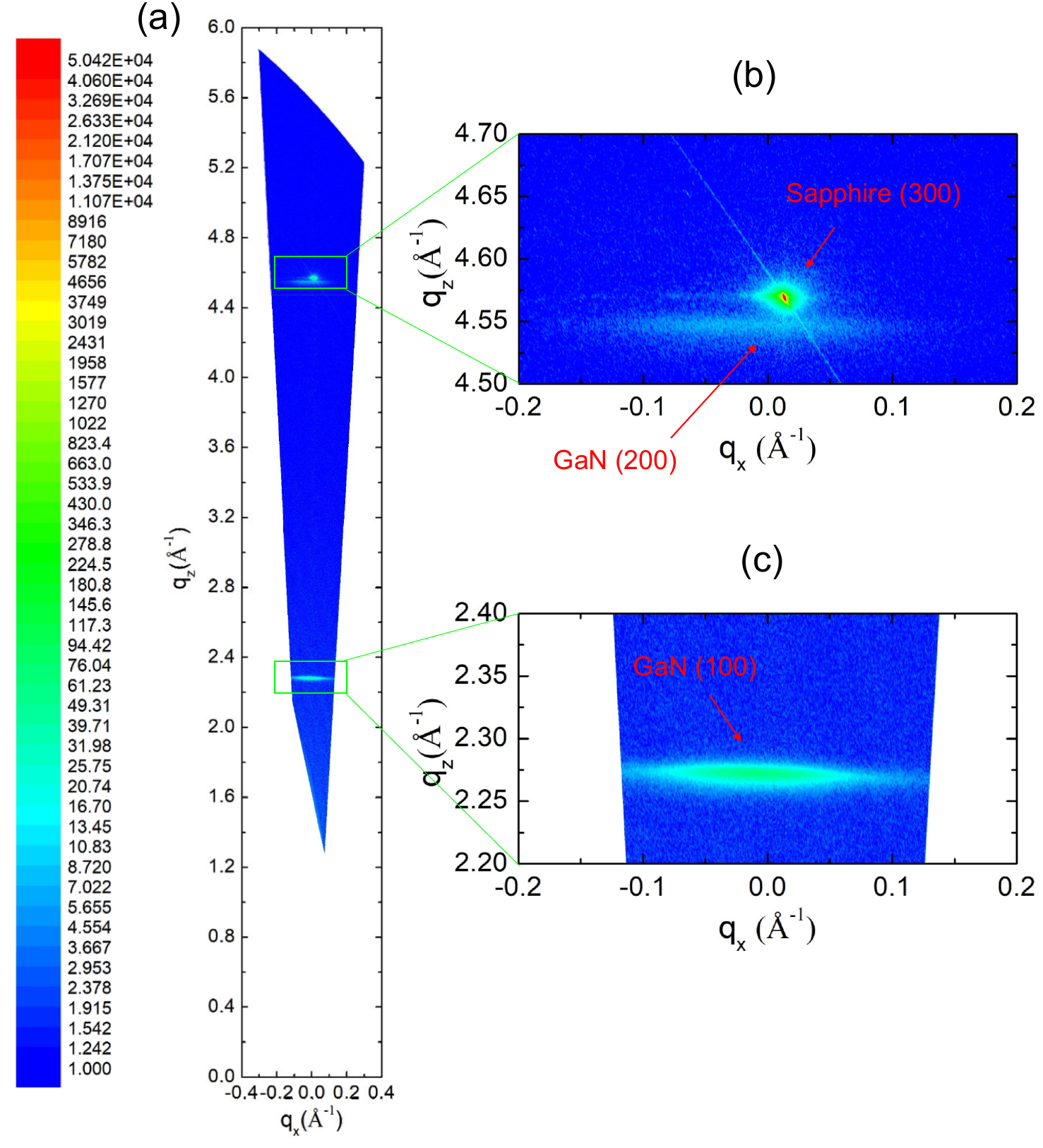}
\caption{(a) The reciprocal space map of X-ray diffraction of GaN grown on a bare $m$-plane sapphire substrate, S1. (b)$\sim$(c) Magnified reciprocal space maps clearly show that (100) Bragg peak at $q_z=2.270$~\AA$^{-1}$ and (200) Bragg peak at $q_z=4.547$~\AA$^{-1}$ are the only Bragg peaks of GaN observed when grown on substrate S1.  Of course, sapphire (300) Bragg peak at $q_z=4.570$~\AA$^{-1}$ is also seen.} 
\label{XRD-100}
\end{figure}

Additionally, we conducted reciprocal space maps (RSMs) of X-ray diffraction for GaN grown on both substrate S1 and S2 to provide rigorous confirmation of the crystallographic orientation of GaN. As depicted in Fig.~\ref{XRD-100}, the RSM for GaN grown on the bare $m$-plane sapphire substrate S1 displays (100) and (200) Bragg peaks, confirming that GaN on this substrate exclusively exhibits (100)-orientation.

\begin{figure}
\includegraphics[width=1.0\columnwidth]{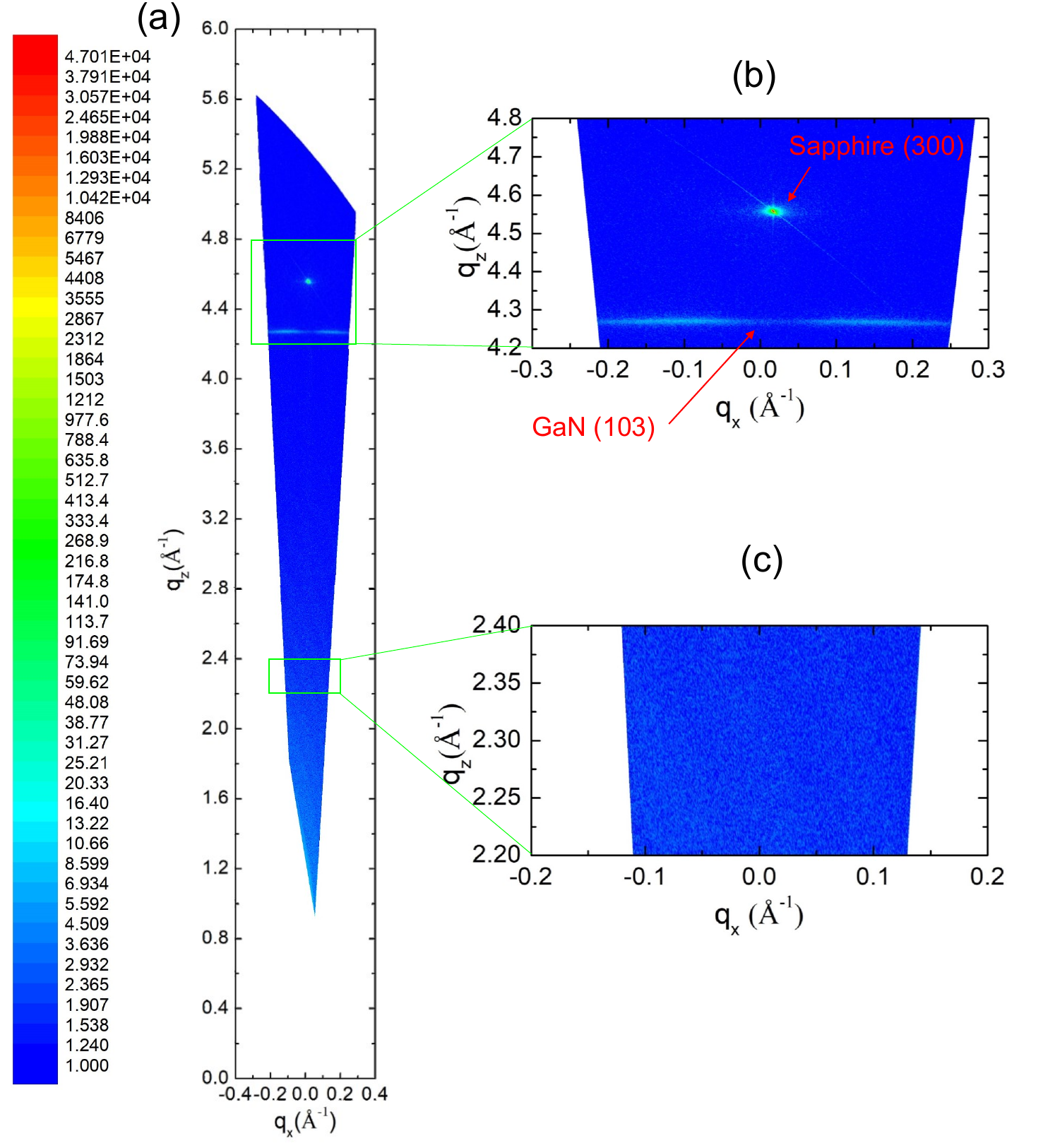}
\caption{The reciprocal space map of X-ray diffraction of GaN grown on substrate S2. (b)$\sim$(c) Magnified reciprocal space maps clearly show that (103) Bragg peak at $q_z=4.270$~\AA$^{-1}$ is the only Bragg peak of GaN observed when grown on substrate S2 while sapphire (300) Bragg peak at $q_z=4.570$~\AA$^{-1}$ is of course observed.  GaN (103) Bragg peak is split because of the slight tilt of twins in opposite directions.\cite{Jue-APL-104-092110}  Note that neither (100) nor (200) GaN Bragg peak is observed when grown on substrate S2.} 
\label{XRD-103}
\end{figure}

Conversely, Fig.~\ref{XRD-103} reveals a different scenario for substrate S2, where the RSM exhibits a (103) Bragg peak that is split into two components. This result indicates that GaN grown on $m$-plane sapphire with a graphene areal fraction less than unity predominantly possesses a (103)-orientation. Note that the apparent splitting of the (103) Bragg peak is attributable to the tilting of two variants in opposite directions, as previously reported\cite{Jue-APL-104-092110}, and is not due to potential fluctuations caused by graphene. The observed tilt angle for each variant is approximately $\sim$1.7\,$^{\circ}$, consistent with the previously reported tilt angle of $\sim$2.0\,$^{\circ}$.\cite{Jue-APL-104-092110}

\begin{figure}
\includegraphics[width=0.8\columnwidth]{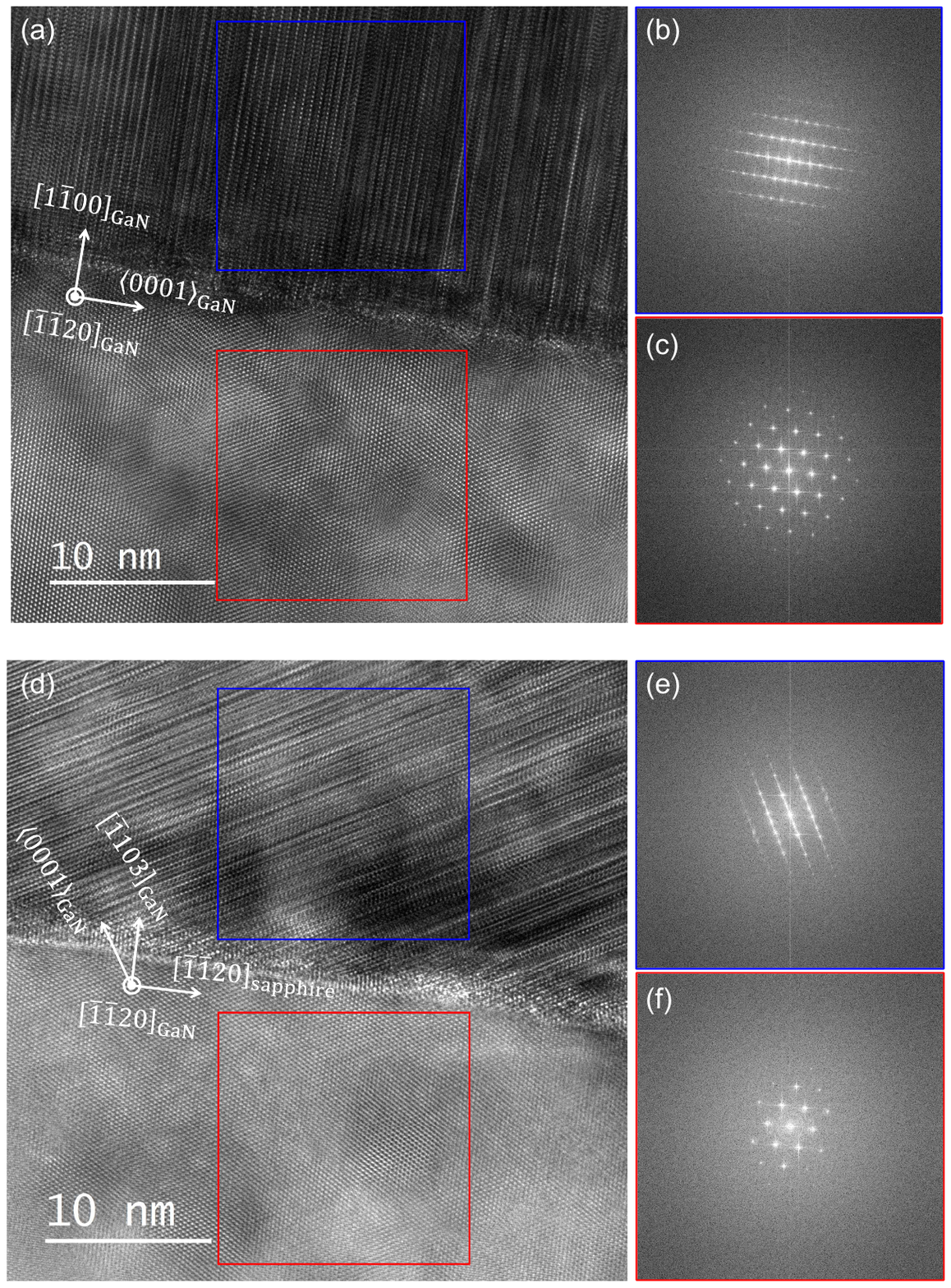}
\caption{High-resolution cross-sectional TEM image of the interfacial region between GaN and $m$-plane sapphire for the case when grown on (a) substrate S1 and (d) substrate S2, respectively.  FFT of the GaN and sapphire regions marked by blue and red squares are shown in (b) and (c) for case I and (e) and (f) for case II, respectively.} 
\label{TEM}
\end{figure} 

The next question we aimed to answer is whether the dissimilarly oriented GaN domains originated from nuclei with differing orientations on the exposed areas of each $m$-plane sapphire substrate. To address this matter, we conducted high-resolution cross-sectional transmission electron microscopy (TEM). As illustrated in Fig.~\ref{TEM}, both (100)- and (103)-oriented GaN were found to be in direct contact with the $m$-plane sapphire substrate, indicating that two dissimilarly oriented GaN domains nucleated under the same growth conditions on two $m$-plane sapphire substrates (one bare (S1) and the other partly covered with graphene (S2)) placed within the same reactor. Notably, the areal fraction of graphene, as shown in the AFM topography, is much less than unity, and graphene is not observed in the high-resolution cross-sectional TEM image. Therefore, these data show that dissimilarly oriented domains can nucleate on the same substrate with different graphene coverages (0 and 0.45, respectively), despite the absence of any influence from the potential supposedly induced by graphene. Consequently, the presence of dissimilarly oriented domains cannot be attributed solely to the modified potential—\textit{i.e.}, the superposed potential from both graphene and $m$-plane sapphire.

All experimental results, including those from SEM, XRD, and TEM, consistently demonstrate differences in the crystallographic orientation of GaN when grown on $m$-plane sapphire, with or without the presence of graphene. Notably, in the case of GaN growth on substrate S2, GaN initially nucleated on the exposed $m$-plane sapphire substrate before laterally growing over graphene, rather than the other way around. TEM shows that GaN can exhibit varying crystallographic orientations when grown on the same $m$-plane sapphire substrate under identical growth conditions, depending on whether the substrate is partly covered by graphene as a mask material.  This cannot be attributed to GaN with varying crystallographic orientations being nucleated on graphene due to a modified substrate potential, as previously suggested.\cite{Du-NL-22-8647} Instead, it arises from GaN with different crystallographic orientations directly nucleating on $m$-plane sapphire.

Orientation disparity on $m$-plane sapphire is not specific to graphene masking. In earlier work on SiO$_2$-patterned $m$-plane sapphire, we observed systematic changes in GaN orientation with opening size and spacing, which we linked to changes in the effective arrival rate at exposed areas supported by DFT trends.\cite{Jue-SR-5-16236} The present results are consistent with the view that, even when nucleation occurs on exposed $m$-plane sapphire, process factors beyond ``mask present vs absent''—including surface conditioning and local flux—can bias orientation within a fixed growth recipe.

These observations show that the crystallographic orientation is not solely determined once the film material and substrate material are selected.  Previously, the crystallographic orientation of GaN when grown on bare $m$-plane sapphire substrates was observed to vary depending on the growth conditions,\cite{Seo-APEX-5-121001} but in this work, we showed that the orientation disparity can be also observed under identical growth conditions on bare and partly graphene-covered $m$-plane sapphire substrates.  In other words, factors such as growth conditions and mask coverage also play a significant role in determining the preferred crystallographic orientation of the material when directly nucleated on the substrate.  This observation counters the assumption made in the previous paper that dissimilar crystallographic orientations cannot result from pinhole-seeded lateral epitaxy.\cite{Du-NL-22-8647}  Of course, there are situations in which mask coverage does not cause orientation disparity of GaN, particularly when grown on $c$-plane sapphire substrates.  The effect of growth conditions and mask coverage on the orientation disparity is not universal and valid in a limited situation.  However, the observation of thru-hole epitaxy responsible for orientation disparity is sufficient enough to serve as counter-example to the role of remote epitaxy exclusively considered responsible for the orientation disparity in the aforementioned paper.\cite{Du-NL-22-8647}

One caveat is that the growth of graphene by CVD requires $m$-plane sapphire exposed to environment gas for several hours, which could possibly modify the surface states of $m$-plane sapphire.\cite{Curiotto-SS-603-2688}  It is also well known that the preferred orientation of GaN grown on $m$-plane sapphire is sensitive to the pretreatments to $m$-plane sapphire.\cite{Paduano-JCG-367-104,Zhang-Micromachines-12-1153}  So, the investigation of preferred orientation of GaN was further carried out by adopting thermally-annealed (S3) and reduced-graphene-oxide(GO)-covered (S4) $m$-plane sapphire substrates.

\subsection{Anneal-only and reduced-GO-covered $m$-plane sapphire}

\begin{figure}
\includegraphics[width=0.8\columnwidth]{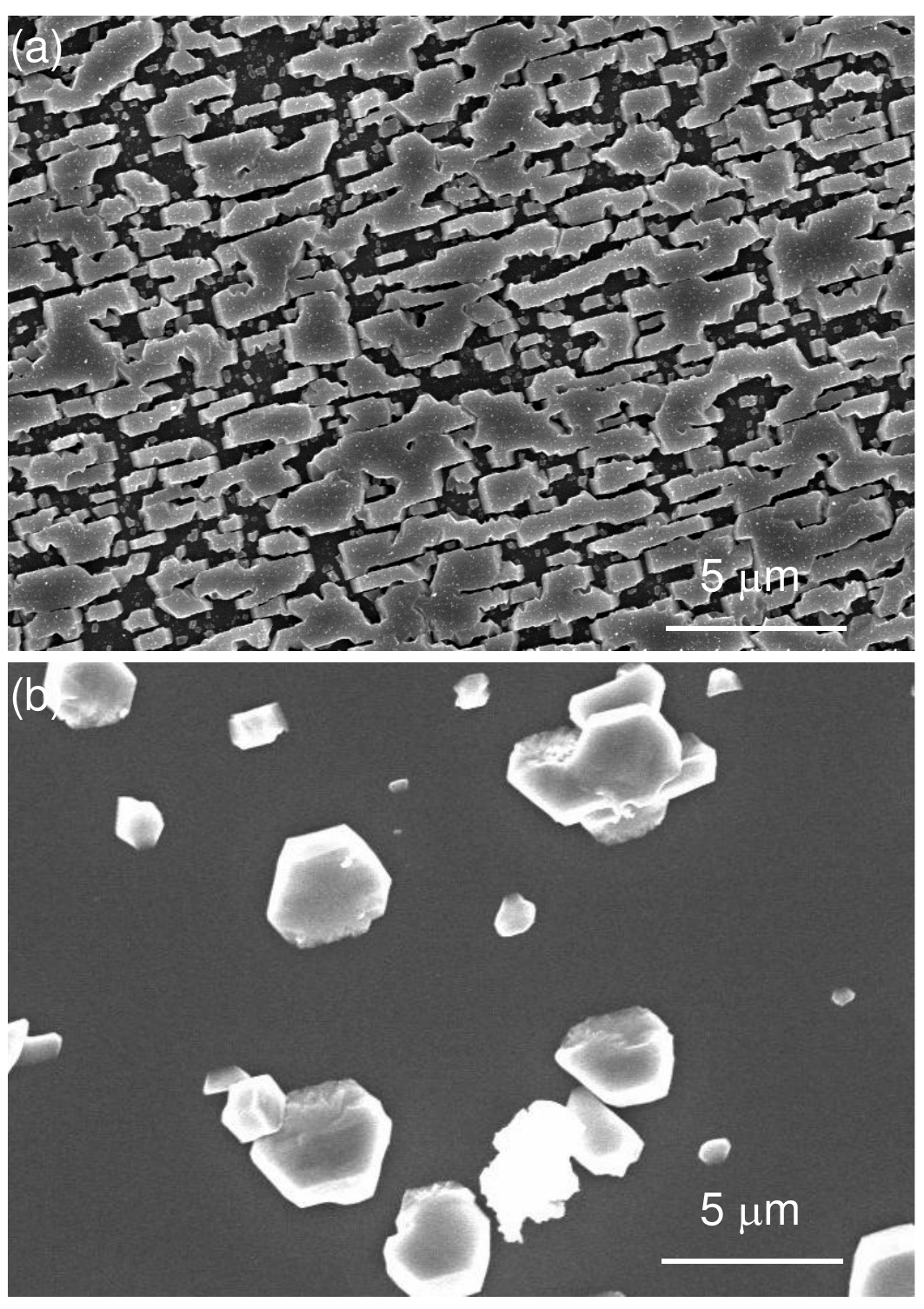}
\caption{SEM images of GaN domains grown on (a) bare $m$-plane sapphire without thermal pre-anneal (S1) and (b) $m$-plane sapphire annealed in Ar/H$_2$ with CH$_4$ off (S3). Both samples were grown simultaneously under identical conditions.  Note that the domain morphology in (b) matches the (103)-oriented case shown in Fig.~\ref{SEM-dissimilar-preferred-orientations-evidence}(c), indicating a (103) preferred orientation. Thus, S1 is (100)-oriented, whereas S3 is (103)-oriented.} 
\label{SEM-annealed}
\end{figure}

Building on the evidence of thru-hole epitaxy above, we isolate substrate pre-treatment effects from mask presence by comparing S1 (bare), S3 (anneal-only, CH$_4$ off), and S4 (rGO, no high-$T$ pre-anneal) within one growth window.  Within one window, we find that high-temperature Ar/H$_2$ pre-anneal primarily sets orientation on $m$-plane sapphire, while the open area fraction of exposed sapphire exerts a secondary bias under THE.  To decouple substrate pre-treatment from mask presence, we prepared two $m$-plane sapphire substrates without any graphene mask: (S1) a bare reference with no pre-treatment, and (S3) a substrate subjected to the same high-$T$ Ar/H$_2$ sequence used before graphene CVD, but with CH$_4$ turned off (i.e., anneal-only). Both were loaded side-by-side and grown in the same run. On the bare reference (S1), GaN domains with the preferred (100) orientation formed, confirming that the baseline window selects (100) on bare $m$-plane sapphire (Fig.~\ref{SEM-annealed}(a)). By contrast, the Ar/H$_2$-annealed substrate (S3) showed a markedly lower nucleation density together with a switch to the (103) preferred orientation (Fig.~\ref{SEM-annealed}(b)). These results indicate that prolonged Ar/H$_2$ annealing—even without CH$_4$—modifies the $m$-plane sapphire surface in a way that alters both nucleation density and preferred orientation.  

\begin{figure}
\includegraphics[width=1.0\columnwidth]{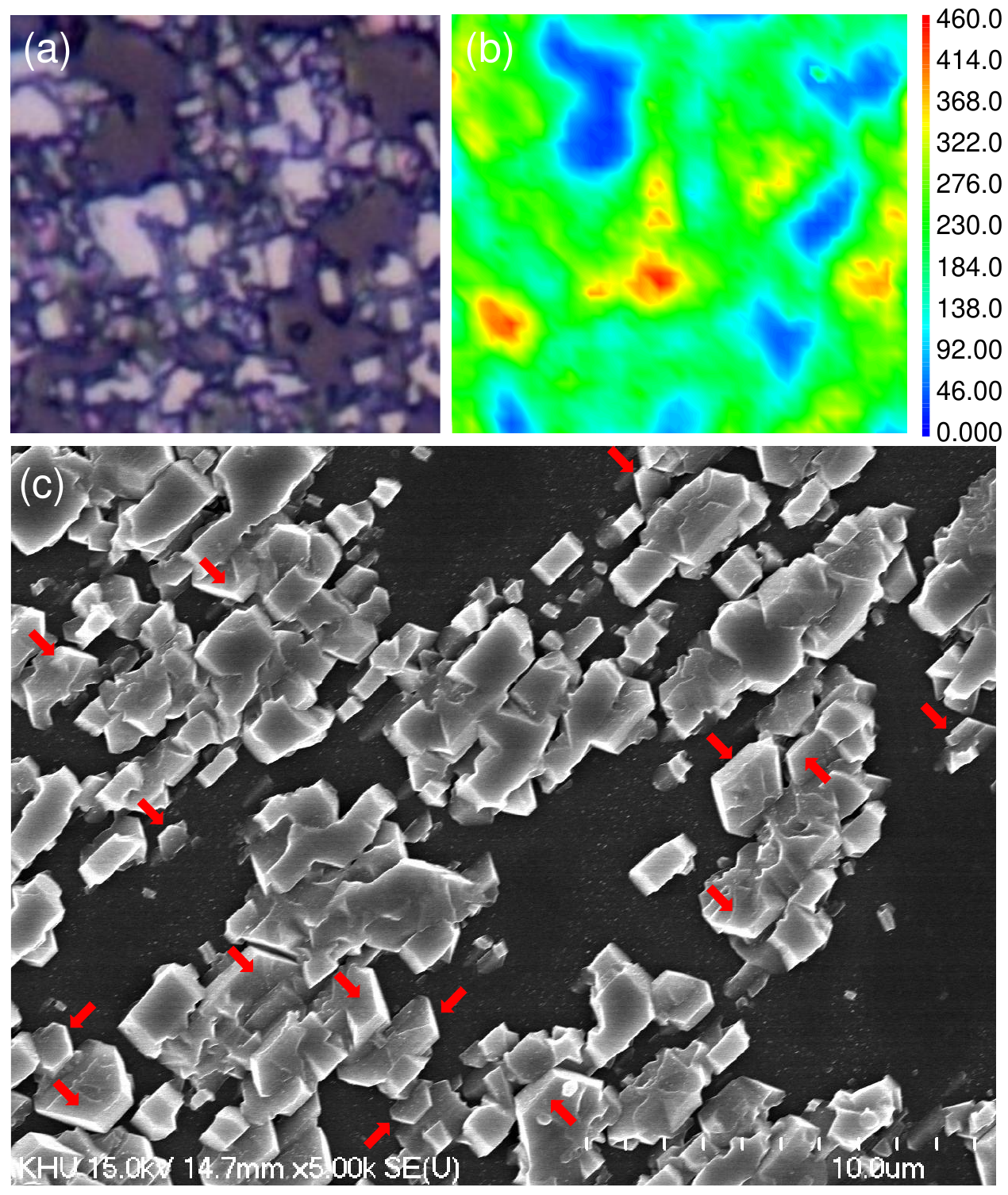}
\caption{(a) Optical micrograph and (b) Raman G-peak ($\sim$1600~cm\(^{-1}\)) intensity map of the S4 sample (GaN grown on spin-coated, reduced GO flakes on $m$-plane sapphire). Panels (a) and (b) are co-registered to the same field of view; in (a), bright regions correspond to GaN domains, whereas dark regions indicate areas without GaN.  (c) SEM image of S4. Most domains are (100)-oriented; (103)-oriented GaN domains are indicated by red arrows.} 
\label{rGO-robustness}
\end{figure}

Previously, spin-coated graphene oxide (GO) flakes reduced to rGO on $c$-plane sapphire were shown to enable thru-hole epitaxy.\cite{Beak-CGD-25-7557} Accordingly, GO flakes were spin-coated onto $m$-plane sapphire and reduced to rGO prior to GaN growth (S4).  Figures~\ref{rGO-robustness}(a,b) show GaN domains on rGO/sapphire; the graphene G band is detected only beneath the GaN islands because the overgrown GaN shields rGO during unload (uncovered rGO oxidizes/vanishes). This co-localization does not imply nucleation on rGO or enhanced nucleation over rGO; rather, GaN nucleated at sapphire exposed through thru-holes and then laterally overgrew the rGO (thru-hole epitaxy). In other words, regions with a higher density of thru-holes nucleate GaN more readily, and those GaN domains subsequently protect the underlying rGO, yielding the observed G-peak signal under GaN but not in rGO-only areas. For clarity, we note that in our prior work on rGO/$c$-plane sapphire the G-peak remained visible even in regions without GaN;\cite{Beak-CGD-25-7557} the difference arises from a milder post-growth handling/unload in that study, whereas the present samples experienced a more oxidative/thermal unload that preferentially removed exposed rGO. This post-growth effect does not affect the nucleation locus or the thru-hole interpretation.  As shown in Fig.~\ref{rGO-robustness}(c), the preferred orientation is predominantly (100), with a minority of (103)-oriented domains. Because S4 received no high-$T$ Ar/H$_2$ pre-anneal, substrate surface modification can be reasonably ruled out; thus, the open-area fraction of exposed sapphire remains a plausible contributor to orientation selection.  


Comparing all cases clarifies the hierarchy: S3 (anneal-only, CH$_4$ off) selects (103), matching S2 (partly CVD-graphene–covered), whereas S4 (rGO on pristine, no high-temperature Ar/H$_2$ pre-anneal) is predominantly (100) with a minority (103). Thus, on $m$-plane sapphire the high-temperature Ar/H$_2$ exposure—i.e., substrate modification—primarily sets the orientation, and mask presence alone does not; the open-area fraction under thru-hole epitaxy provides only a secondary bias. Taken together, S1/S3/S4 support this ordering.

\section{Conclusion}
Under one growth window, GaN on $m$-plane sapphire selects (100) on bare $m$-plane sapphire (S1) and (103) on partly graphene-covered $m$-plane sapphire (S2). Cross-sectional TEM shows that nuclei in S2 contact exposed sapphire (thru-hole), not graphene—so orientation disparity can occur even when nucleation is on the substrate, without invoking a superposed potential from the 2D layer. Comparing S1, S2/S3 (“graphene-grown”/anneal-only), and S4 (rGO on pristine) indicates that high-temperature Ar/H$_2$ pre-treatment primarily sets orientation, while mask presence alone does not; open-area fraction can provide a secondary bias within the thru-hole regime. This work shows one concrete example in which orientation by itself does not determine the growth mechanism; mechanism assignment should control and decouple mask continuity and substrate pre-treatment within the same process window.

\section{acknowledgement}
This work was supported by the National Research Foundation of Korea(NRF) grant funded by the Korea government (MSIT) (RS-2021-NR060087, RS-2023-00240724) and through Korea Basic Science Institute (National research Facilities and Equipment Center) grant (2021R1A6C101A437) funded by the Ministry of Education.



%

\end{document}